\input epsf
\documentstyle[12pt]{article}

        {\newpage
        \setcounter{page}{1}}

\renewcommand{\title}[1]{%
        {\begin{center}
        \Large\bf #1
        \end{center}}
        \vskip .3in}

\renewcommand{\author}[1]{%
        {\begin{center}
        #1
        \end{center}}}

\renewcommand{\abstract}[1]{%
        \begin{center}%
        {\vspace{1em}\vspace{0pt}\bf Abstract}%
        \end{center}%
        \noindent #1}

\renewcommand{\date}[1]{%
        \begin{center}%
        #1%
        \end{center}}

        {\end{thebibliography}}

\makeatletter
        \@addtoreset{equation}{section}%
\makeatother

\newcommand{\eqn}[1]{\label{eq:#1}}
\newcommand{\refeq}[1]{(\ref{eq:#1})}
\newcommand{\eq}{eq.~\refeq}
\newcommand{\eqs}{eqs.~\refeq}

\newcommand{\beq}{\begin{eqnarray}}
\newcommand{\eeq}{\end{eqnarray}}

\newcommand{\eu}{\epsilon_u}
\newcommand{\ed}{\epsilon_d}
\newcommand{\es}{\epsilon_s}
\newcommand{\eN}{\epsilon_N}

\newcommand{\semidirect}{\mathbin{\hbox{\hskip2pt\vrule height 5.2pt
depth -.3pt width .25pt \hskip-2pt$\times$}}}
\newcommand{\mybar}[1]%
        {\kern 0.8pt\overline{\kern -0.8pt#1\kern -0.8pt}\kern 0.8pt}

\newcommand{\drawsquare}[2]{\hbox{%
\rule{#2pt}{#1pt}\hskip-#2pt
\rule{#1pt}{#2pt}\hskip-#1pt
\rule[#1pt]{#1pt}{#2pt}}\rule[#1pt]{#2pt}{#2pt}\hskip-#2pt
\rule{#2pt}{#1pt}}

\newcommand{\Yfund}{\raisebox{-.5pt}{\drawsquare{6.5}{0.4}}}
\newcommand{\Yasymm}{\raisebox{-3.5pt}{\drawsquare{6.5}{0.4}}\hskip-6.9pt%
        \raisebox{3pt}{\drawsquare{6.5}{0.4}}}

\newcommand{\jref}[4]{{\it #1} {\bf #2}, #3 (#4)}
\newcommand{\NPB}[3]{\jref{Nucl.\ Phys.}{B#1}{#2}{#3}}
\newcommand{\PLB}[3]{\jref{Phys.\ Lett.}{#1B}{#2}{#3}}
\newcommand{\PR}[3]{\jref{Phys.\ Rep.}{#1}{#2}{#3}}
\newcommand{\PRD}[3]{\jref{Phys.\ Rev.}{D#1}{#2}{#3}}
\newcommand{\PRL}[3]{\jref{Phys.\ Rev.\ Lett.}{#1}{#2}{#3}}

\begin{document}

\begin{titlepage}
\begin{center}
{\hbox to\hsize{hep-ph/9705411 \hfill  DOE/ER/40561-325-INT97-00-171}}
{\hbox to\hsize{               \hfill  UW/PT-97-12}}
{\hbox to\hsize{               \hfill  BUHEP-97-15}}

\bigskip
\bigskip
\bigskip
\vskip.2in

{\Large \bf Flavor from Strongly Coupled Supersymmetry} \\

\bigskip
\bigskip
\bigskip
\vskip.2in

{\bf David B. Kaplan$^a$, Fran\c{c}ois Lepeintre$^a$ and  Martin 
Schmaltz$^b$}\\

\vskip.2in

{ \small \sl $^a$
Institute for Nuclear Theory,   Box 351550

University of Washington,

Seattle, WA 98195-1550
}

\smallskip

{\tt dbkaplan@phys.washington.edu, francois@u.washington.edu}

\bigskip
\bigskip

{\small \sl $^b$ Department of Physics

Boston University

Boston, MA 02215, USA }

\smallskip

{\tt schmaltz@abel.bu.edu}

\vspace{1.5cm}
{\bf Abstract}\\
\end{center}

\bigskip

Strongly coupled supersymmetric theories can  give rise to
composite quarks and leptons at low energy.  We show that the internal
structure of these particles can explain the origin of three generations
and provide a qualitative understanding of  mass ratios and mixing
angles between the different flavors of fermions,
all within a renormalizable theory.
The main point of the paper is to  show how fermion masses and mixing angles
can result from  a ``dual'' Frogatt-Nielsen mechanism: fields neutral under
$SU(3)\times SU(2)\times U(1)$
which carry flavor quantum numbers are confined within quarks and
leptons, and from their perturbative interactions arises the observed flavor
structure.

\bigskip

\end{titlepage}

\section{Introduction}

The large hierarchy between the electroweak and Planck scales suggests  that
electroweak symmetry breaking is  associated with nonperturbative
physics in analogy with the hierarchy between
$\Lambda_{QCD}/M_P$. Various theoretical arguments also
indicate
that the world may be supersymmetric at short distances (see, for example,
\cite{susyref}). While supersymmetry (SUSY) protects the electroweak scale
from radiative corrections and associates it with the SUSY breaking scale,
it does not by itself explain the origin of the large hierarchy.
Thus it is often suggested that a  SUSY theory with nonperturbative
dynamics lies behind the Standard Model (SM).

In much of the literature it is assumed that the nonperturbative
SUSY  physics lies in a hidden sector, decoupled from Standard Model
particles  \cite{susyref}. It seems to us a rather strong assumption that
nonperturbative physics exists at short distance solely to fix the electroweak
scale (or the SUSY breaking scale), without affecting any other low energy
observables. In this paper we speculate
that strong interactions are not only responsible for the electroweak
hierarchy but also for a substructure for quarks and leptons which
could explain family replication, mass hierarchies and flavor mixing.
Our motivation is partly opportunistic --- there have been significant
advances in the past several years in the understanding of strongly
interacting SUSY theories \cite{Seib,exactres}, and one can now construct
models of composite quarks and leptons \cite{copapers} with a degree of 
theoretical
confidence impossible until recently.  The development of new theoretical
tools begs application to the old problem of flavor.  As we discuss at length
below, compositeness can provide a simple explanation for why families
are exact replicas of each other, as far as $SU(3)\times SU(2)\times U(1)$
gauge charges are concerned, while being distinguished by their masses and
mixing angles.  An analogy can be made between the SM generations and
nuclear isotopes. Consider, for example, the three isotopes of hydrogen: they
each have the same chemistry  yet have dramatically
different masses --- a fact simply  understood once it is realized that the
nucleus is composite, and that the three isotopes each contain a single proton
but varying numbers of neutrons.  Similarly, quarks and leptons could be
bound states of both charged and neutral constituents, with different numbers
or types of neutral constituents for the different generations.  The nature of
these composites will be determined by the underlying strong interactions, and
the interactions of the ``neutrons'' will largely determine the flavor 
structure
observed at low energies.

In this paper we show how a strongly coupled SUSY theory can realize this
paradigm for the origin of three families, particle masses, and flavor mixing.
We begin by discussing the replication of families.  We then devise a new
mechanism
for flavor structure along the lines of the isotope analogy discussed above. We
then describe several renormalizable models in which all quarks and leptons are
composite, and which reproduce qualitatively the flavor structure we observe in
the Standard Model. We conclude with speculations on future directions along
the
lines proposed here.

\section{Family replication}

The first issue to be addressed is what sort of strongly coupled SUSY gauge
theory to consider. A minimum requirement is that the theory must have
composite particles in its spectrum which transform non-trivially under
a sufficiently large symmetry group to contain $SU(3)\times SU(2) \times U(1)$.
(In this work, we simplify our task by not trying
to simultaneously explain flavor physics and dynamical SUSY breaking.)\
As discussed at length in \cite{CSS}, the properties of $N=1$ supersymmetric
gauge theories without a tree level superpotential are largely determined by
the number
\beq
\chi \equiv \sum_j \mu_j - \mu(G)\ ,
\eeq
where $\mu_j$ and $\mu(G)$ are the Dynkin indices for the matter and adjoint
representations respectively of the gauge group $G$. Normalizing $\mu=1$ for
the fundamental representation, $\chi$ is even.
Theories with $\mu(G)<\chi<0$ have runaway vacua and no groundstate or
break supersymmetry.
Theories with $\chi> 2$ have moduli spaces of inequivalent vacua with massless
gauge bosons at the origin; for many of these theories dual descriptions are
known.
Of the theories with $\chi=0$ or $\chi=2$ many are known to confine;
in the case of $\chi=0$ confinement always occurs with a quantum deformed
moduli space which breaks  chiral symmetry, whereas confining $\chi=2$ theories
have unbroken chiral symmetry at the origin of moduli space and dynamically
generate a superpotential for the confined fields.

The $\chi<0$ theories do not appear promising
to us; although the runaway vacua may be stabilized by superpotential terms,
the minimum  will typically be characterized by vacuum expectation values
(VEVs) for fields that break the global symmetries, which we wish to
preserve. The $\chi > 2$ theories are interesting --- one can imagine that the
Standard Model is the  dual of some strongly coupled theory, with the quarks
and leptons being the magnetic degrees of freedom, and the SM  gauge group
being the dual gauge group after possible partial spontaneous symmetry
breaking.  However, working backward from the SM to find a dual typically leads
to enormous gauge groups  and we have yet to see or think of a clever approach
of this type.
The confining $\chi=2$ theories  are particularly interesting for several
reasons.  First, they exhibit confinement and possess large unbroken
global symmetries.  Secondly, these ``s-confining'' theories,
have been completely classified in \cite{CSS}, and their dynamically generated
superpotentials can be constructed straightforwardly.  We focus on these
theories because they are the best understood and possess the properties we
desire.
(Confining $\chi=0$ theories might also be interesting even though 
--- or because \cite{CHM} --- the quantum deformed moduli space forces partial
chiral symmetry breaking. Their properties can
often be derived from the s-confining $\chi=2$ theories by giving a large
mass to a flavor so that it decouples.)

A particularly intriguing example of an $s$-confining theory is an $Sp(2N)$
gauge theory with six fundamentals
$Q$ and an antisymmetric tensor $A$ \cite{Cho,oursp}.  The theory
has an $SU(6)\times U(1)$ global symmetry, as well as an $R$ symmetry.  The
confined description involves the  $Sp(2N)$ neutral fields
\beq
T_m &=& {\rm Tr} A^m, \ m=2,3,\ldots,N,\nonumber\\
M_n &=& Q A^n Q, \ n=0,1,2,\ldots,N-1\ ,
\eeq
where $Sp(2N)$ indices are contracted with the appropriate metric, which can be
taken to be $J = i\sigma_2\times 1_N$ \footnote{The number  of families equals
$N$ because the matrix $A J$ has eigenvalues which come in pairs;  thus it
satisfies the square root of its characteristic equation, implying that
$A^{N+1}$ may be expressed in terms of  lower powers of $A$.}.  The quantum
numbers of the fields
are:

\beq
 \begin{array}{c|c|ccc}
    & Sp(2 N) & SU(6) & U(1) & U(1)_R \\[.1in] \hline
&&&&\\[-.1in]
  A & \Yasymm   & 1     & -3   & 0 \\
  Q & \Yfund    & \Yfund& N-1& \frac{1}{3} \\[.1in] \hline \hline&&&&\\[-.1in]
  T_m    & & 1       & -3m  & 0 \\
  M_n   &  & \Yasymm & 2 (N-1) -3 n & \frac{2}{3}
 \end{array}
\eqn{sp2n}
\eeq
This model has a number of desirable features, and will be the ``workhorse''
of all the explicit  models discussed below. If weak gauge interactions are
embedded in the $SU(6)$ symmetry of this model, then there is a replication of
``families'' of $M$ fields~\cite{oursp}. Furthermore,
in spite of having $N$ families, the  family symmetry of the model is not
$U(N)$,  but only
$U(1)$. Family replication arises because the $A$ field only carries this
global $U(1)$ charge, and so the SM gauge charges of a composite particle are
independent of the number of $A$ fields it contains.  Breaking this $U(1)$
flavor symmetry will allow us to generate flavor in a manner analogous to the
Froggatt-Nielsen mechanism. The model realizes the isotope paradigm of the
introduction, with  the $(QQ)$ and $A$ fields  playing the roles  of the proton
and neutron respectively.

\section{A new  mechanism for generating texture}

If all Yukawa interactions in the Standard Model were to vanish there would be
a global $U(3)^5$ chiral flavor symmetry.  The real Yukawa couplings
explicitly
break this chiral symmetry down to $U(1)_B\times U(1)_L$, but in a hierarchical
manner, as apparent from the hierarchy of observed  fermion masses and mixing
angles.
Effective low energy models of the symmetry breaking can be constructed: one
makes assumptions about what subgroup  $H\in U(3)^5$  is the approximate
chiral symmetry at short distances,  and then introduces spurions which break
$H$ down to $U(1)_B\times U(1)_L$.  The advantage of such models is that the
large mass  hierarchies we observe may be explained qualitatively in terms of
several parameters of  order $1/10$, however the analysis is never unique due
to the paucity of information about the Yukawa matrices available to us
experimentally.  The analysis is further complicated  in supersymmetric
theories, where the physics of flavor and of SUSY breaking may be intertwined,
as in general squark and slepton masses are sensitive to both.

To proceed, one must go beyond the effective description of flavor in terms of
spurions, and construct models for the origin and communication of flavor
symmetry  breaking.  It is possible that the origins of flavor lie above the
Planck scale, or that  the breaking of flavor symmetries is due to ``Planck
slop'' --- nonrenormalizable operators  suppressed by powers of $M_P$.
However, without a renormalizable theory of flavor, one  gains little insight
beyond that obtained from the spurion analysis.

A perturbative framework for generating flavor texture was proposed in
\cite{FN} by Froggatt and Nielsen (FN).  The proposal is that at short distance
the only  global flavor symmetry consists of $U(1)$ groups.  Quarks and leptons
are coupled to  heavy, vector-like fields $V$, which in turn  couple to a field
$A$ with strength  $g$, such that the product $gA$ can be consistently assigned
$U(1)$ charges. (Either the $A$  is  charged, or $g$ is a $U(1)$ spurion, or
both.)\  When the heavy $V$ fields are  integrated out of the
theory, nonrenormalizable operators involving the quarks and leptons are
generated, involving powers of $g A/M_V$.  Below $M_V$, the $A$ field acquires
a VEV, and  the quantity $\epsilon\equiv g\langle A\rangle/M_V$ serves as the
flavor spurion  with which the Yukawa couplings are constructed.  Various
$\epsilon$'s can be  assigned charges under the $U(1)$ flavor symmetries,
and the different elements of  the Yukawa matrices are constrained by the quark
and lepton charge assignments to be  proportional to different powers of
$\epsilon$ \cite{abelflav}.  This framework can be generalized to  incorporate 
nonabelian
symmetries \cite{nonabelflav}, both discrete and continuous.

The drawback of the FN approach is that whereas  the 54 real parameters of the
SM Yukawa matrices (in a particular basis) are traded for typically far fewer
charges and VEVs, the charge assignments and symmetry breaking patterns tend to
look quite {\it ad hoc} and  little  insight is gained  into the origins of
flavor.

The mechanism for generating flavor structure that we propose here is similar
to the FN model, except that instead of having the $A$ field get an expectation
value, we have it carry strong interactions that cause it to be confined within
the quarks and leptons.  Thus the spurion
characterizing the  flavor hierarchies is not $g\langle A\rangle/M_V $, but
rather $\epsilon =g\Lambda/M_V$, where $\Lambda$ is the confinement scale of
the quark and lepton constituents.   Since the
$A$ field is in a confined phase, instead of a Higgs phase as in the FN
scenario, our mechanism is in some sense dual to the FN mechanism. The
advantage of this approach is twofold:  the FN  charges of the quarks and
leptons are now set by the number of $A$ constituents,  determined by dynamics
rather than fiat.  Also, the two mass scales appearing in  $\epsilon$, $M_V$
and $\Lambda$ are not entirely independent since typically the strong group
will run
much faster after the $V$ fields are integrated out, so that $\Lambda$ will
not be far below $M_V$.

In the next section we give an example of a toy model which realizes the
features we have been discussing --- the low energy theory consists of
composite fields  for which both family replication and hierarchical Yukawa
interactions arise as a  consequence of the internal structure of the
composites.

\section{A toy model with $SU(3)^3$ symmetry}

We now construct a model which illustrates the flavor mechanism discussed
above.  It is far from realistic --- leptons end up being the heaviest
particles, $u$ and $d$ quarks have proportional mass matrices, and  the CKM
matrix is trivial. Nevertheless, it nicely illustrates how nontrivial texture
can arise dynamically in a renormalizable model where all SM particles are
composite.  In the subsequent section we will discuss more realistic models.

\subsection{Fundamental fields and interactions}

Yukawa interactions have the generic form $LHR$, where $L$ and $R$ correspond
to the left- and right-handed fermions and $H$ is the Higgs.  If all three
fields are composite, such an interaction might be generated nonperturbatively
due to instantons, in which case, following the power counting scheme of ref.
\cite{EffSUSY}, the Yukawa coupling would be $\sim 4\pi$.  Thus  only the top
Yukawa interaction
can be due purely to strong dynamics, and  we assume that the weaker Yukawa
interactions arise from some interplay between strong and perturbative
interactions.  In the models we construct, $L$ and $R$ (with the exception of
the top, in one example) are composed of constituents bound by different strong
forces, and are able to interact only due to perturbative interactions.  In the
toy model we present here, we assume that the Higgs is composite as well, and
thus that the strong interaction corresponds to a semisimple gauge group with
three factors.

 Since we wish to explain the existence of three families, we take the strong
group to be $Sp(6)^3 = Sp(6)_L \times Sp(6)_H \times Sp(6)_R$. To create
composite SM fields we include  a single antisymmetric  tensor and six
fundamentals for each $Sp(6)$ factor, as in \eq{sp2n}.  Without any
perturbative  superpotential
added to  the theory, the model possesses a global $SU(6)^3$ symmetry, in which
we can embed ``trinification'', the $SU(3)^3 \semidirect Z_3$ GUT introduced
in~%
\cite{trinification} where the $Z_3$ symmetry cyclicly permutes the three
$SU(3)$
group factors. A SM family is embedded in the ``trinified" representation
$(3,\mybar 3,1) \oplus (1,3,\mybar 3) \oplus (\mybar 3,1,3)$.
We take  the fundamental ``preons'' to transform under $Sp(6)^3\times SU(3)^3$
as~%
\footnote{From here on we use the notation that strongly
coupled
fundamental fields are lowercase, while composite fields are in capitals.}:

\beq
\begin{array}{c|ccc|ccc}
{\rm preon} & Sp(6)_{L} & Sp(6)_{H} & Sp(6)_{R}& SU(3)_1 & SU(3)_2 & SU(3)_3
 \\[.1in] \hline &&&&&& \\[-.1in]
a_1 & \Yasymm &1 & 1  & 1  &1   & 1  \\
p_1 & \Yfund  &1  &1   & 3&1   & 1  \\
q_1 & \Yfund  &1  & 1  &1   & \mybar 3 & 1   \\[.1in]
\hline &&&&&&\\[-.1in]
a_2 &1  & \Yasymm & 1  & 1  & 1  & 1   \\
p_2 &1  & \Yfund & 1  & 1 & 3 &1   \\
q_2 &1  & \Yfund &1   &1  &1   & \mybar 3 \\[.1in]
\hline &&&&&&\\[-.1in]
a_3 &1  &1  & \Yasymm & 1  &1   &1   \\
p_3 &1  &1  & \Yfund &1   &1   & 3 \\
q_3 &1  &1  & \Yfund & \mybar 3 &1   &1   \\
\end{array}
\eqn{preonsi}
\eeq
\medskip

Each of the $Sp(6)$ factors gives rise to composite fields as discussed in
\S2:
there are the $T_2= a^2$ and $T_3=a^3$ fields, which are neutral under
$SU(3)^3$, as well as the  composite fields shown below:

\beq
\begin{array}{c|ccc}
{\rm composite} & SU(3)_1 & SU(3)_2 & SU(3)_3  \\[.1in]
 \hline \\[-.1in]
 \Phi^{(1)} =p_1 q_1 & 3 & \mybar 3 & 1 \\
 \Phi^{(2)} =p_2 q_2 & 1 & 3 & \mybar{3} \\
 \Phi^{(3)} =p_3 q_3 & \mybar{3} & 1 & 3 \\[.1in]
\hline \\[-.1in]
X =q_3 q_3 & 3 &1 & 1 \\
\mybar X =p_1 p_1 & \mybar 3 &1 & 1 \\
Y =q_1 q_1 & 1 &3 & 1 \\
\mybar Y =p_2 p_2 & 1 &\mybar 3 &1  \\
Z =q_2 q_2 & 1 &1 & 3 \\
\mybar Z =p_3 p_3 & 1 & 1 &\mybar 3 \\

\end{array}
\eqn{compositesi}
\eeq

\medskip

\noindent
For simplicity we have only listed the family with no $a$ constituents; there
are in fact three families. For example, the three families of  $\Phi^{(1)}$
fields are $(p_1 q_1)$, $(p_1 a_1 q_1)$, and $(p_1 a_1^2 q_1)$.  In this model,
the fields with the most $a$ constituents will be the lightest, and so these
composites correspond to the third, second, and first family respectively.

The SM gauge group is contained in $SU(3)^3$  by identifying
$SU(3)_{c}=SU(3)_1$,while  embedding $SU(2)_{w}\subset SU(3)_2$ and
$U(1)_Y\subset
SU(3)_2\times SU(3)_3$. With this embedding, the $\Phi$ fields decompose as
\beq
 \Phi^{(1)} & \to & Q\oplus G \nonumber\\
 \Phi^{(2)} & \to & L\oplus \mybar E\oplus \mybar N\oplus H_u\oplus H_d\oplus
S \\
 \Phi^{(3)} & \to & \mybar U\oplus \mybar D\oplus \mybar G\nonumber
\eeq
where $Q,\mybar U,\mybar D,L,\mybar E,H_{u,d}$ are fields with SM quantum
numbers,
$\mybar N$ is a right-handed neutrino, $G$ is a $(3,1)_{-1/3}$ exotic diquark,
and $S$ is a singlet.  A virtue of this model is that all exotic fields are in
real
representations of the SM and can in principle acquire large masses.

We now add a perturbative superpotential so that the three sets of composite
fields can
interact with each other.  The purpose of this exercise is to generate a
superpotential in the low energy theory that includes a Yukawa interaction
between composites  of the three strong groups, which exhibits a hierarchical
structure among the three  families. To generate the desired interaction we
introduce three heavy fields $v_i$, each of which transform as fundamentals
under two of the  strong $Sp(6)$ groups:

\beq
\begin{array}{c|ccc|ccc}
{\rm heavy\ field} & Sp(6)_{L} & Sp(6)_{H} & Sp(6)_{R}& SU(3)_1 & SU(3)_2 &
SU(3)_3  \\[.1in] \hline &&&&&& \\[-.1in]
v_1 &1 & \Yfund & \Yfund & 1 & 1 & 1\\
v_2 & \Yfund  &1 & \Yfund & 1 & 1 & 1\\
v_3 & \Yfund & \Yfund & 1 & 1 & 1 & 1  \\
\end{array}
\eqn{vi}
\eeq
\medskip

With these fields we write down a nongeneric, renormalizable superpotential
consistent with an $Sp(6)^3 \times SU(3)^3 \semidirect Z_3$
symmetry:
\beq
\eqn{vcoup}
W&=& \sum_{i=1,2,3}  \frac{1}{2}{\cal M}_i v_i^2 +  \beta v_1 v_2 v_3 + \gamma
p_{i-1}v_i q_{i+1}\nonumber \\
{\cal M}_i &\equiv& \left( \mu-\alpha_+ a_{i+1} - \alpha_- a_{i-1}\right)\ .
\eeq
where the subscripts are integers modulo 3, $\mu$ is the common mass of
the $v$ fields, and $\alpha_\pm$, $\beta$ and
$\gamma$ are coupling constants. This superpotential has the necessary
couplings for the three sectors to communicate. Note that the $a v^2$ couplings
break the $U(1)^3$ family symmetry that counts the $a$ fields, thus allowing
mixing between families.

\subsection{The effective superpotential}

We integrate out the $v$ fields at their mass  scale  $\mu$, assumed to be
above the confinement scale $\Lambda$, and expand the effective superpotential
in powers of the coupling $\beta$. The result at lowest order arising from
the contribution in fig.~1 is
\begin{figure}[t]
\centerline{\epsfxsize=3 in \epsfbox{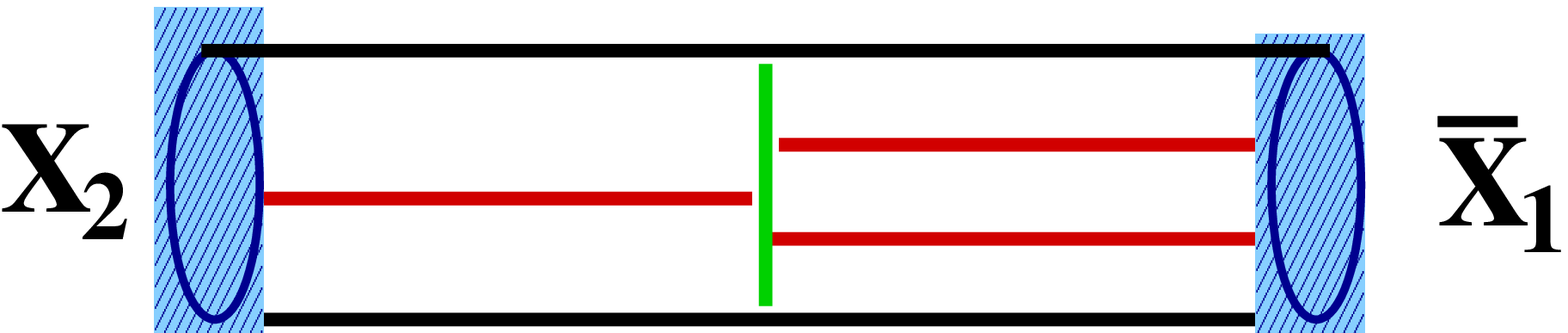}}
\vskip .2in
\noindent
Fig 1. {\it Contribution to the effective superpotential at order $\beta^0$
from
integrating out the massive ${\rm v}$ fields.  Exterior lines are the ${\rm
p}$ and ${\rm q}$
preons;
internal lines are ${\rm v}$ propagators, from which ${\rm a}$ preons are
emitted.  The
${\rm p}$,
${\rm q}$ and ${\rm a}$ preons later confine, and below the confinement scale
this
contribution to $W_{eff}$ is a mass term, where the number of ${\rm a}$
constituents
determines the family numbers of the composites.}
\vskip .1in
\end{figure}

\beq
\eqn{WbelowM}
W_0 = -{\gamma^2\over 2} \sum_i {(p_{i+1}q_{i-1})^2 / {\cal M}_i}\ .
\eeq
At the scale $\Lambda$, the theory confines and the operator
\refeq{WbelowM} gets mapped onto  a mass term for composite fields.

By expanding the propagators $1/{\cal M}_i$ in \eq{WbelowM} in powers of
each of the  $a_i$ fields up to second order, we can compute the effective
superpotential in terms of confined fields\footnote{In the following
expressions we set to zero the $T$ moduli,  the singlet fields composed
entirely of $a$ constituents. It is not difficult to include them if one
wishes.}. If we denote the families with subscripts $n=1,2,3$ such that
$n=1$ corresponds to the lightest family (maximum number of $a$ constituents)
while $n=3$ corresponds to the heaviest family (no $a$ constituents), then   at
order $\beta^0$ we find that the three families of real exotics $X,\mybar
X,Y,\mybar Y,Z,\mybar Z$ acquire hierarchical masses
\beq
\eqn{W0belowL}
W_0\to \left({\Lambda^2\over \mu}\right)M_{mn} \left( \mybar X_m X_n
+ \mybar Y_m Y_n + \mybar Z_m Z_n \right)
\eeq
where the mass matrix $M$ is given by ($r,s=0,1,2$)
\beq
M_{3-r,3-s}&=&\left({\Lambda\over \mu}\right)^{r+s}
 {1 \over r! s!}{\partial^r_x}{\partial^s_y}
\left( 1 -\alpha_+ x - \alpha_- y\right)^{-1}\bigl.\bigr\vert_{x=y=0}
 = {(r+s)!\over r! s!}\epsilon_+^{r}\epsilon_-^{s}\nonumber\\ \nonumber \\
&=&\left(\matrix{
  6  \epsilon_-^2\epsilon_+^2 & 3  \epsilon_-\epsilon_+^2 &\epsilon_+^2\cr
 3  \epsilon_-^2\epsilon_+ & 2   \epsilon_-\epsilon_+ &\epsilon_+ \cr
 \epsilon_-^2  & \epsilon_- &1 \cr
 }\right)
\eqn{mass}
\eeq
with the definition
\beq
\epsilon_\pm \equiv \left({\Lambda\over \mu}\right) \alpha_\pm\ .
\eeq

While we are able to reliably calculate the effective superpotential
we cannot determine the K\"ahler potential and with it the wave function
renormalization for the composite fields. However, in the absence of
small parameters in the strong dynamics we can assume that the relative
wave function renormalization of the different composites differs at most
by ${\cal O}(1)$ factors and does not wash out the predicted hierarchy.

At order $\beta$, the effective superpotential arises from the graph in
fig.~2, which yields
%
\begin{figure}[t]
\centerline{\epsfxsize=3 in \epsfbox{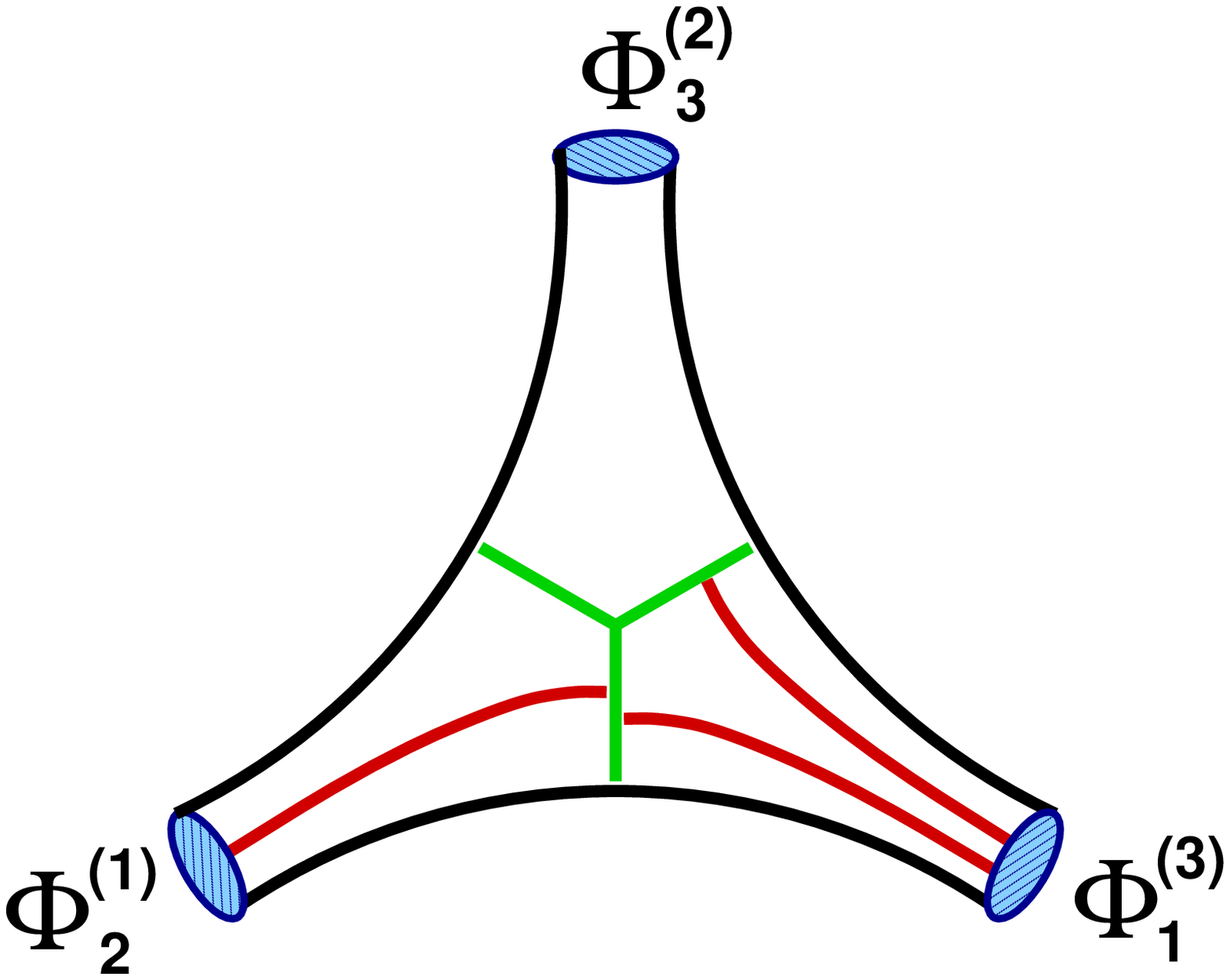}}
\vskip.2in
\noindent
Fig 2. {\it Contribution to the effective superpotential at order $\beta$
from integrating out the massive ${\rm v}$ fields. Below the confinement
scale,
these contributions become trilinear interactions between composites.}
\vskip .2in
\end{figure}
\beq
W_1&=& -\beta\gamma^3 {(p_1 q_1)(p_2q_2)(p_3q_3) / \left({\cal M}_1{\cal
M}_2{\cal M}_3\right)} \nonumber \\
&\to& \beta\left({\Lambda\over \mu}\right)^3 y_{rst} \Phi^{(2)}_r
\Phi^{(1)}_s \Phi^{(3)}_t\ .
\eqn{wone}
\eeq
where the three families of chiral composite fields $ \Phi^{(i)}$ interact via
the
 Yukawa coupling
\beq
\begin {array}{ll}
\eqn{yukdef}
y_{3-r,3-s,3-t}=\left({\Lambda\over \mu}\right)^{r+s+t}
{1\over r!s!t!}\quad\times&\\[.1in]
\qquad{\partial^r_x}{\partial^s_y}{\partial^t_z}
\left((1 -\alpha_+ x - \alpha_- y)
(1 -\alpha_+ y - \alpha_- z)
(1 -\alpha_+ z - \alpha_- x)\right)^{-1}\bigl.\mid_{x=y=z=0}& .
\end{array}
\eeq
For example, the Yukawa matrix $ y_{3mn}$ has entries
\beq
\eqn{yzero}
 y_{3mn}= \left(\matrix{
(\epsilon_++\epsilon_- )^4+3
\epsilon_+^2\epsilon_-^2 & (\epsilon_++\epsilon_- )^3
+ \epsilon_+^2\epsilon_- &(\epsilon_++\epsilon_- )^2- \epsilon_+\epsilon_- \cr
(\epsilon_++\epsilon_- )^3+ \epsilon_+\epsilon_-^2&
 (\epsilon_++\epsilon_- )^2+ \epsilon_+\epsilon_- &
(\epsilon_++\epsilon_- )\cr
(\epsilon_++\epsilon_- )^2- \epsilon_+\epsilon_- &(\epsilon_++\epsilon_-)&1\cr
}\right)
\eqn{toyyuk}
\eeq
We need not continue the $\beta$ expansion any further, since at higher order
in
$\beta$ the effective superpotential contains at least four pairs of $p$ and
$q$ preons, which translates into a nonrenormalizable interaction below the
confinement scale.

These Yukawa interactions \eq{toyyuk} contribute to $u$ and $d$ quark
masses, provided that a pair of the three families of $H_{u,d}$ can be
identified with the MSSM Higgs fields. However, since
 the composite Higgs and lepton fields arise from the same  strong group
$Sp(6)_H$,
interactions among themselves arise purely from nonperturbative physics, and
not from $v$ exchange.
Thus, to address the question of Higgs and lepton masses, one must consider
nonperturbative contributions to the superpotential.  The nonperturbative
superpotential has been worked out for $Sp(6)$ in~\cite{Cho,oursp}.  In
the notation of table \refeq{sp2n}, where subscripts denote the number of $a$
constituents, and the composites have not yet been rescaled to mass dimension
1, it is given by

\beq
W_{dyn}= \frac{1}{\Lambda^7} \Big( \frac{1}{3} T_2^2 M_0^3
     + \frac{1}{2} T_3 M_1 M_0^2 - \frac{1}{2} T_2 M_0^2 M_2 +
      \frac{1}{4} M_0 M_2^2 + \frac{1}{4} M_1^2 M_2 \Big) \ .
\eqn{Sp6superpot}
\eeq
The $M$ fields are antisymmetric tensors of $SU(6)$, while $T$ fields are
singlets; subscripts denote the number of $a$ constituents.
Under the $SU(3)^3$
decomposition \refeq{compositesi}, $M^3 \to
(\Phi^{(2)})^3 + \mybar Y \Phi^{(2)} Y$ for the
$Sp(6)_H$ composites, etc.  The SM content of the three $\Phi^{(2)}_a$ fields
may be written as a $3\times 3$ matrix
\beq
\Phi^{(2)}_a = \left(\matrix{L&H_d&H_u\cr S&\mybar N&\mybar E\cr}\right)_a
\eeq
where the dynamical interaction $[\Phi^{(2)}]^3$ has indices contracted with
two $\epsilon$ tensors as in a determinant. Among the interactions one
finds\footnote{Above in eq. (\refeq{Sp6superpot}) the subscripts denote
the number of $a$ constituents; for the remainder of the paper subscripts
will denote family number, defined as ($3-$number of $a$ constituents).}
\beq
\Phi^{(2)}_a\Phi^{(2)}_b\Phi^{(2)}_c = S_a H_{u,b} H_{d,c} + H_{d,a} L_b \mybar
E_c +  H_{u,a} L_b \mybar N_c +\ldots
\eeq
Comparing the above expression with $W_{dyn}$ in \eq{Sp6superpot}, it is
interesting to note that in the vacuum where the modulus $S_1$ has a VEV
while other moduli vanish, the three families of
$H_{u,d}$ fields have a dynamically generated mass matrix which is rank 2:
\beq
M^2_H \propto \langle S_2\rangle   \left(\matrix{0&1&0\cr 
1&0&0\cr0&0&0\cr}\right).
\eeq
Thus only the Higgs fields without an $a$ constituent survive to the weak scale
in this vacuum.

While the above mechanism is an interesting way to ensure that a single pair of
Higgs fields survive to low energy, it has unwelcome phenomenological
consequences.  One is  that the superpotential contains a global $SU(2)_R$
symmetry relating $\mybar U \leftrightarrow \mybar D$ and $\mybar N
\leftrightarrow
\mybar E$.
Thus $u$ and $d$ quarks both have Yukawa interactions given by \refeq{toyyuk},
their mass ratios are equal and mixing angles vanish.  Furthermore, one sees
that
the dynamically generated lepton mass matrices are rank 1  with one family
being very massive (heavier than the top quark, as the $\beta
(\Lambda/\mu)^3 $ suppression appearing in  \eq{wone} is absent) while the
other two lepton families are massless.

\subsection{Comments}

The model we have presented here is certainly no candidate for beyond the SM
physics;  however, it does provide an example of how a renormalizable
field theory can give rise to three composite families of SM fields with
nontrivial flavor texture.  The structure of the mass and Yukawa matrices of
\eqs{mass},\refeq{toyyuk} is sufficiently complex that  one can imagine
more sophisticated models based on our mechanism being at least in qualitative
agreement with the flavor structure of the SM. This is the subject of the next
section.

We conclude this section with a comment on the factor $\Lambda/\mu$ that
appears in \eq{wone}, and which phenomenologically cannot be be
allowed to be very small. Recall that $\Lambda$ characterizes the confinement
scale, while $\mu$ is the mass of the $v$ fields that communicate between the
Higgs and the left- and right-handed fields.  One might worry about tuning a
perturbative coupling to be close to a nonperturbative scale.  However, note
that the fundamental parameters of the theory in the ultraviolet are not $\mu$ 
and
$\Lambda$, but rather $\mu$ and $\Lambda_{UV}$, where the latter is the scale
that determines the running of the $Sp(6)$ interactions above the scale $\mu$.
The relation between $\Lambda_{UV}$ and the confining scale $\Lambda$ is easily
estimated from the one-loop beta functions, and one finds in the present theory
that
\beq
\left({\Lambda\over \mu}\right) =\left({\Lambda_{UV}\over \mu}\right)^{1/7}\ .
\eeq
Thus, for example, on could have $\Lambda_{UV}/\mu$ vary from $10^{-3}$ to
$10^{-2}$, while $\Lambda/\mu$ only varies between  $0.4$ and $0.5$. The lack
of fine tuning reflects the fact that the $Sp(6)$ groups are nearly
asymptotically flat with the $v$ fields included, but confine quickly once the
$v$'s are integrated out.

\section{More realistic models}

In this section we construct two models based on the $Sp(6)$ confining theory,
which succeed in reproducing qualitatively much of the flavor structure seen in
the SM.  The purpose of these models is to show that a renormalizable, strongly
coupled theory can give rise to  nontrivial CKM angles and fermion mass ratios.
The structure of model 1 is similar to the toy model of the previous
section but with a cure for most of the shortcomings of the simpler toy. It
possesses the interesting feature that the top Yukawa coupling is generated
nonperturbatively from $Sp(6)$ instanton interactions.  However, the model
cannot reproduce realistic masses and mixing angles.
Model  2  is of slightly different structure and succeeds in fitting  all the
masses and
mixing angles of the Standard Model.
Both models are rather complicated, and are intended to serve as existence
proofs rather than as paragons of beauty.

\subsection{$Sp(6)^3$ with composite Higgs fields}

The main drawbacks of the toy model of the previous section were that (i) there
were no CKM angles; (ii) the top quark Yukawa coupling was of order $\beta
(\Lambda/\mu)^3$, which is likely to be too small; and (iii) the lepton masses
were generated dynamically, and were therefore either too large or zero.  The
first model we examine is similar to the  example of the previous section, but
is designed to correct the three major deficiencies. In particular,  the up
Yukawa matrix receives both dynamical and
perturbative contributions, allowing the top to be very heavy. The down type
couplings are only generated perturbatively, and so they are naturally light.
As up and down sectors are now treated differently, nontrivial mixing angles
 and dissimilar mass ratios are possible.
Finally, lepton masses are also generated perturbatively, but at higher order
than the down quarks, so that they are naturally lighter.  The price for these
successes is that the model has less symmetry and is more complicated. It makes
gauge coupling unification look mysterious, and it predicts incorrect 
relations, such as $m_\mu/m_e \sim 1700$, and
$V_{us}\sim V_{cb}$.

The strong group is taken to be $Sp(6)^3$, and we take the same preons as in
the previous example, with the exception that there is no $SU(3)^3$ symmetry,
and the preons have different $U(1)$ assignments.  These $U(1)$ assignments
are unique under the requirement that (i) it is possible for the top quark
Yukawa coupling to arise nonperturbatively, and (ii) that all charged exotics
are real under $SU(3)\times SU(2)\times U(1)$ so that they can in principle be
heavy.

\beq
\begin{array}{c|ccc|ccrr}
{\rm preon} & Sp(6)_{L} & Sp(6)_{H} & Sp(6)_R & SU(3) & SU(2) & U(1)_Y &
U(1)_{B-L} \\[.1in]
\hline&&&&&&&\\[-.1in]
a_1 & \Yasymm & 1 & 1 & 1 & 1 & 0 & 0 \\
t_1 & \Yfund & 1 & 1 & 3 & 1 & -1/3 & -1/6 \\
d_1 & \Yfund & 1 & 1 & 1 & 2 & 1/2 & 1/2 \\
s_1 & \Yfund & 1 & 1 & 1 & 1 & 0 & -1/2 \\[.1in]
\hline&&&&&&&\\[-.1in]
a_2 & 1 & \Yasymm & 1 & 1 & 1 & 0 & 0 \\
d_2 & 1 & \Yfund & 1 & 1 & 2 & -1/2 & -1/2 \\
s_{21} & 1 & \Yfund & 1 & 1 & 1 & 0 & -1/2 \\
s_{22} & 1 & \Yfund & 1 & 1 & 1 & 0 & 1/2 \\
s'_{22} & 1 & \Yfund & 1 & 1 & 1 & 0 & 1/2 \\
s_{23} & 1 & \Yfund & 1 & 1 & 1 & 1 & 1/2 \\[.1in]
 \hline&&&&&&&\\[-.1in]
a_3 & 1 & 1 & \Yasymm & 1 & 1 & 0 & 0 \\
t_3 & 1 & 1 & \Yfund & \mybar 3 & 1 & 1/3 & 1/6 \\
s_{31} & 1 & 1 & \Yfund & 1 & 1 & 0 & -1/2 \\
s_{32} & 1 & 1 & \Yfund & 1 & 1 & 0 & 1/2 \\
s_{33} & 1 & 1 & \Yfund & 1 & 1 & -1 & -1/2 \\
\end{array}
\eqn{preonsii}
\eeq

\noindent
The first subscript on the preon fields designates under which $Sp(6)$ group
they transform. As before, ``a'' designates an $Sp(6)$ antisymmetric tensor,
while ``s'', ``d'', and ``t'' label $Sp(6)$ fundamentals which transform as
singlets, doublets, and triplets respectively under $SU(3)\times SU(2)$.

As before, the $Sp(6)^3$ groups confine (assumed for simplicity to occur at
the same scale) and the composite fields transform as the
three families of the SM (with right-handed neutrinos), plus exotic states
which are all real under
$SU(3)\times SU(2)\times U(1)$.  Their quantum numbers are

\beq
\begin{array}{ll|ccrr}
\multicolumn{2}{c|}{\rm composite} & SU(3) & SU(2)& U(1)_Y & U(1)_{B-L}  
\\[.1in]
 \hline&&&&&\\[-.1in]
\mybar U^{(1)} \!\!\!\!&= t_1 t_1 & \mybar 3 & 1 & -2/3 & -1/3 \\
\mybar E^{(1)} \!\!\!\!&= d_1 d_1 & 1 & 1 & 1 & 1 \\
Q \!\!\!\!&= t_1 d_1 & 3 & 2 & 1/6 & 1/3 \\
G \!\!\!\!&= t_1 s_1 & 3 & 1 & -1/3 & -2/3 \\
H_u^{(1)} \!\!\!\!&= d_1 s_1 & 1 & 2 & 1/2 & 0 \\[.1in]
 \hline&&&&&\\[-.1in]
E^{(2)} \!\!\!\!&= d_2 d_2 & 1 & 1 & -1 & -1 \\
L \!\!\!\!&= d_2 s_{21} & 1 & 2 & -1/2 & -1 \\
H_d^{(2)} \!\!\!\!&= d_2 s_{22} & 1 & 2 & -1/2 & 0 \\
{H'}_d^{(2)} \!\!\!\!&= d_2 s'_{22} & 1 & 2 & -1/2 & 0 \\
H_u^{(2)} \!\!\!\!&= d_2 s_{23} & 1 & 2 & 1/2 & 0 \\
S^{(2)} \!\!\!\!&= s_{21} s_{22} & 1 & 1 & 0 & 0 \\
{S'}^{(2)} \!\!\!\!&= s_{21} s'_{22} & 1 & 1 & 0 & 0 \\
\phi_+ \!\!\!\!&= s_{21} s_{23} & 1 & 1 & 1 & 0 \\
\mybar N\!\!\!\!&= s_{22} s'_{22} & 1 & 1 & 0 & 1 \\
\mybar E^{(2)} \!\!\!\!&= s_{22} s_{23} & 1 & 1 & 1 & 1 \\
 {\mybar E'}^{(2)} \!\!\!\!&= s'_{22} s_{23} & 1 & 1 & 1 & 1 
\\[.1in]
 \hline&&&&&\\[-.1in]
U \!\!\!\!&= t_3 t_3 & 3 & 1 & 2/3 & 1/3 \\
\mybar D \!\!\!\!&= t_3 s_{31} & \mybar 3 & 1 & 1/3 & -1/3 \\
\mybar G \!\!\!\!&= t_3 s_{32} & \mybar 3 & 1 & 1/3 & 2/3 \\
\mybar U^{(3)} \!\!\!\!&= t_3 s_{33} & \mybar 3 & 1 & -2/3 & -1/3 \\
S^{(3)} \!\!\!\!&= s_{31} s_{32} & 1 & 1 & 0 & 0 \\
E^{(3)} \!\!\!\!&= s_{31} s_{33} & 1 & 1 & -1 & -1 \\
\phi_- \!\!\!\!&= s_{32} s_{33} & 1 & 1 & -1 & 0 \\
\end{array}
\eqn{compositesii}
\eeq

\noindent
The charge assignments have been made so that the first set of preons yields
the composites $Q$, $\mybar U^{(1)}$ and $H_u^{(1)}$,  which have the right 
quantum
numbers to generate a large dynamical top quark Yukawa coupling.

As before, we will add massive $v$ fields to the theory, which will generate
perturbative preon interactions which will in turn become SM Yukawa
interactions below the confinement scale $\Lambda$. Before specifying exactly
what $v$ fields are needed, we sketch out how the various composites get their
masses.

\subsubsection{The $d$ quarks}

Since $Q$,  $H_d/H'_d$ , and $\mybar D$ fields arise from the three 
different
$Sp(6)$ groups,
Yukawa interactions for the $d$ quarks can be
generated via the interaction in fig.~2, as  in the model of the
previous
section.  This will require  three $v$ fields, which we denote $v_{1,2,3}$. The
$d$ Yukawa coupling will then arise from expanding
$1/({\cal M}_1{\cal M}_2{\cal M}_3)$, where ${\cal M}_i$ is the $a$ dependent
mass  of $v_i$, and thus is of the form \refeq{yukdef}, \refeq{yzero}.

\subsubsection{The $u$ quarks}

There are six $\mybar U$ fields in the theory (the three families of $\mybar
U^{(1)}$ and $\mybar U^{(3)}$);  three can get large masses by pairing up with 
the
three $U$ fields. The mass matrix
takes the form
\beq
M_U\sim \bordermatrix{&\bar U^{(1)} &\bar U^{(3)}\cr U& 1/{\cal M}_2 & 
4\pi\langle
S^{(3)}\rangle\cr}\ .
\eeq
The factor of $4\pi$ follows from the power counting arguments
in ref. \cite{EffSUSY}. The coupling between $U$ and
$\mybar U^{(1)}$ arises through the exchange of the
$v_2$ field proportional to $1/{\cal M}_2$ and resembles the mass matrix
$M_{mn}$ in \eq{mass}. The $S^{(3)}$ dependent coupling between $U$ and
$\mybar U^{(3)}$ arises from the nonperturbative potential \refeq{Sp6superpot} 
as
discussed in \S4.2, and its family structure is
\beq
\langle S^{(3)}\rangle \propto \left(\matrix{
S^{(3)}_3& S^{(3)}_2&S^{(3)}_1 \cr
 S^{(3)}_2& S^{(3)}_1 & 0\cr
S^{(3)}_1 & 0 & 0\cr}\right)\ .
\eeq

Since $1/{\cal M}_2$ is rank 3, all three families of $U$ aquire masses. As it
has the hierarchical structure seen in \eq{mass}, the $\mybar U^{(1)}$ field
that couples most strongly to $U$ is the one with no $a_1$ constituents.  By
choosing only $S^{(3)}_1$ to get a VEV, $\langle S^{(3)}\rangle$ is rank 2,  
and
by adjusting its size relative
to  $1/{\cal M}_2$, one can arrange to have the $U$ quarks pair in such a way
that the massless $\mybar U$ quarks include one which is primarily $\mybar 
U^{(1)}$
 (the top), one which is entirely $\mybar U^{(3)}$ (the up), and one which is
mostly $\mybar U^{(3)}$ (the charm). The $\mybar U^{(3)}$ components can 
couple to an
$H_u^{(2)}$ field through a perturbative diagram as in fig.~2, while the $\bar
U^{(1)}$ components can couple nonperturbatively to $H_u^{(1)}$. Therefore we 
must
have some linear combination of $H_u^{(1)}$ and $H_u^{(2)}$ develop a VEV at
the electroweak scale.

\subsubsection{The $H_{u,d}$ doublets}

There are a total of six $H_{u,d}$ pairs in this theory, and we will assume
that only one pair remains light down to the weak scale.  Without loss of
generality, we can take the light down Higgs to be in one of the three families
of  $H_d^{(2)}$ (as opposed to ${H'}_d^{(2)}$).  From the previous 
discussion, we see
that the light $H_u$ must have components in both $H_u^{(1)}$ (to give mass to 
the
top) and in $H_u^{(2)}$ (to give mass to the up and charm). The Higgs mass 
matrix
allows both  $H_d^{(2)}$ and ${H'}_d^{(2)}$ to couple  perturbatively to 
$H_u^{(1)}$, and
nonperturbatively to $H_u^{(2)}$.  We can get the required mass pattern if we 
take
the Higgs mass matrix to look like
\beq
\eqn{mhiggs}
M_H\sim \bordermatrix{
&H_d^{(2)} &{H'}_d^{(2)}\cr
H_u^{(1)}& 0  & 1/{\cal M}_h\cr
H_u^{(2)} & 4\pi\langle {S'}^{(2)}\rangle_2 & 4\pi\langle 
S^{(2)}\rangle_3\cr}
\eeq
where $ 1/{\cal M}_h$ is once again the sort of matrix in \eq{mass} due
to the exchange of a new $v_h$ field, and the subscripts on the $S$ VEVs
signify
the rank.
Evidently, there is a single massless down-type Higgs which is $H_d^{(2)}$ 
(third family), and therefore there is a single massless up-type 
Higgs
field.  The latter contains components both in $H_u^{(1)}$ (primarily first 
family)
and in $H_u^{(2)}$, as desired.

\begin{figure}[t]
\centerline{\epsfxsize=2.5 in \epsfbox{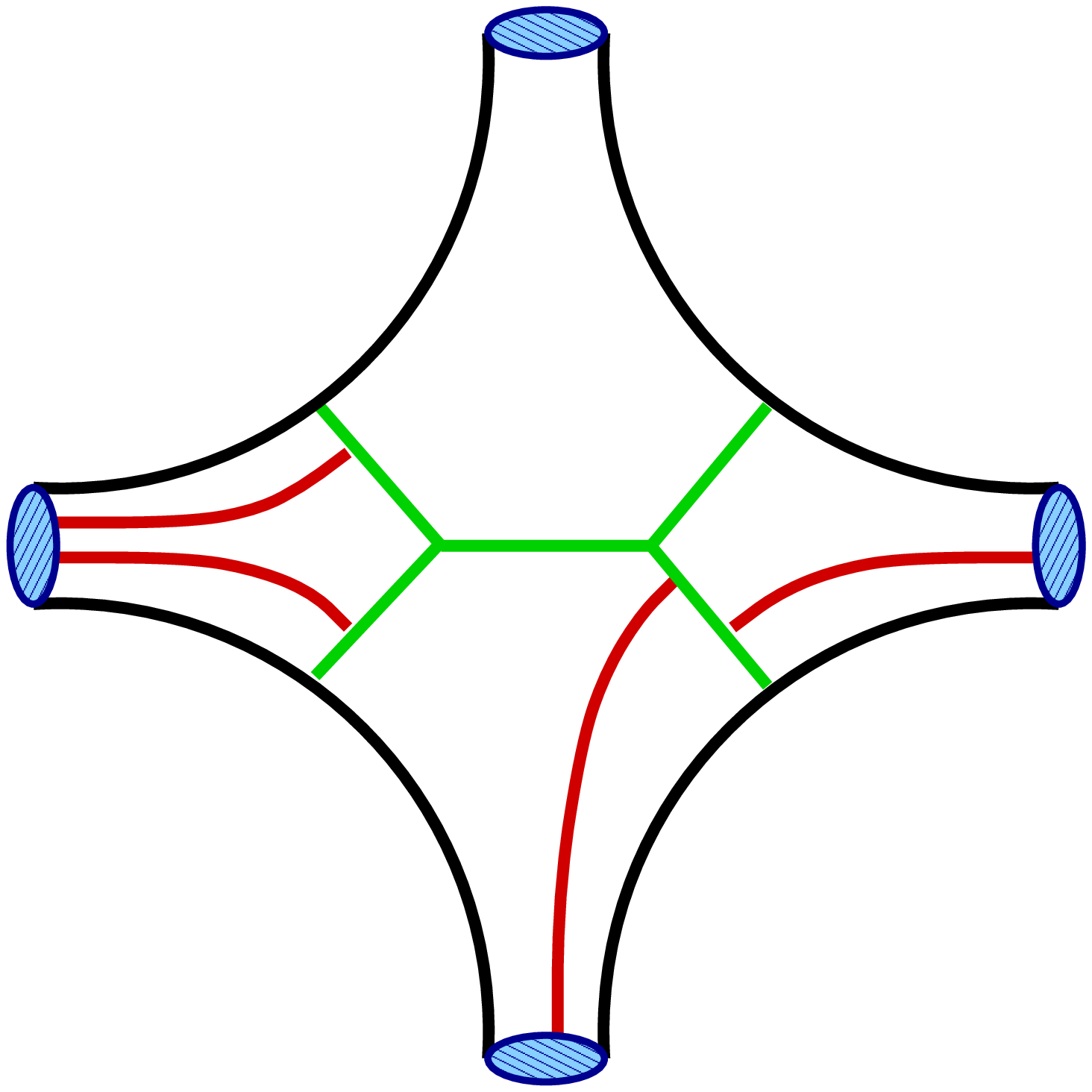}}
\vskip.2in
\noindent
Fig 3. {\it Contribution to the effective superpotential at order $\beta^2$
from
integrating out the massive $v$ fields. Below the confinement scale, these
contributions
become quadrilinear interactions between composites.}
\vskip .2in
\end{figure}
The fact that a pair of Higgs fields remains massless relies on the
$H_u^{(1)}-H_d^{(2)}$ element
in the  matrix \refeq{mhiggs} remaining zero to all orders in perturbation
theory.
A potential problem might arise from ${\cal O}(\beta^2)$ contributions to
$W_{eff}$
shown in fig.~3, which could yield the effective operator
$H_u^{(1)}H_d^{(2)}S^{(3)}{S'}^{(2)}$, for
example, where the two singlets $S^{(3)}$ and ${S'}^{(2)}$ get VEVs.  
However, it is
possible to
avoid these contributions by choosing the couplings of the heavy $v$ fields
appropriately,
so that the zero in \refeq{mhiggs} is preserved at all orders in $W_{eff}$.

\subsubsection{Charged leptons}

The charged lepton mass matrix takes the form
\beq
\eqn{mlepton}
M_E\sim \bordermatrix{
&\bar E^{(2)} & E^{(3)}\cr
\mybar E^{(1)}& 1/{\cal M}_3+1/{\cal M}_h & 0\cr
\mybar E^{(2)}& 4\pi\langle {S'}^{(2)}\rangle_2 & 1/{\cal M}_1\cr
{\mybar  E'}_2& 4\pi\langle S^{(2)}\rangle_3 & 1/{\cal M}_E\cr}
\eeq
where $ 1/{\cal M}_1$,  $1/{\cal M}_3$, $1/{\cal M}_h $, and $1/{\cal M}_E$
arise from exchange of $v_1$, $v_3$, $v_h$, and a new field $v_E$
respectively.  The ranks of the nonperturbative contributions $\langle
S^{(2)}\rangle$, $\langle {S'}^{(2)}\rangle$ were fixed when considering 
the Higgs
matrix above.

The massless $\mybar E$ fields will be family dependent combinations of
$\mybar E^{(1)}$, $\mybar E^{(2)}$, and $\mybar E'^{(2)}$.  Computing the 
Yukawa couplings
to the light Higgs $H_{d,0}^{(2)}$ of these fields is complicated, as the
coupling to $L H_d^{(2)}$ is nonperturbative ($\sim 4\pi$) for $ {\mybar 
E'}^{(2)}$,
zero for $\mybar E^{(2)}$, and perturbative at order $\beta^2$ for $\mybar 
E^{(1)}$
via the graph  pictured in fig.~3.  The $1/{\cal M}_E$ term was introduced in
\eq{mlepton} so that the leptons would not contain much $ {\mybar 
E'}^{(2)}$, which
would lead to excessively large lepton masses.

\subsubsection{Neutrinos}
As it stands, the model predicts that
neutrino masses are similar to quark masses. The situation can be remedied
by giving the right handed neutrinos a large
mass so that the light neutrinos are predominantly left-handed with small
Majorana masses given by the see-saw formula
\beq
M_\nu=- [Y^l]^T M^{-1}_N Y^l \left<H_u\right>^2\ .
\eeq
where $M_N$ is the Majorana mass matrix of the right handed neutrinos.
Since the right handed neutrinos are composite as well their masses
must arise from preon dynamics. A suitable mechanism involves adding
an extra massive field $w_N$ with lepton number $1$ which transforms
as an antisymmetric tensor of the $Sp(6)_H$ group.  We include the
following superpotential couplings
\beq
W_N=\frac{1}{2}(\mu-\alpha_N a_2) w_N^2  + w_N s_{22} s'_{22} \ .
\eeq
Note that the mass term for $w_N$ breaks lepton number by two units as required
for generating a Majorana mass for neutrinos.
Integrating out $w_N$ yields a graph as in fig.~1.  Expanding in $\alpha_N$,
and matching onto
confined fields gives the Majorana mass matrix for right-handed neutrinos
\beq
\eqn{mnr}
M_N \sim {\Lambda^2\over\mu} \left(\matrix{
        6\eN^4 & 3\eN^3 & \eN^2 \cr
       3\eN^3 & 2 \eN^2 & \eN\cr
       \eN^2& \eN & 1  \cr  }\right)\ , \quad
\eN = \alpha_N \frac{\Lambda}{\mu} \ .
\eeq
The resulting masses for the left-handed neutrinos are easily determined
because the dynamically generated Dirac mass matrix for the neutrinos has
only one non-zero entry in the (1,1) component. Thus the electron
neutrino is massive with a mass that is inversely proportional to the
compositeness scale $m_{\nu_e} \sim \langle H_u \rangle^2/\Lambda$, and
the other two neutrino species are massless.

\subsubsection{Numerics}

A summary of the massive fields needed to generate the Yukawa interactions
in this model are as follows:

\beq
\begin{array}{c|ccc}
{\rm field} & Sp(6)_{L} & Sp(6)_{H} & Sp(6)_{R}  \\[.1in]\hline&&&\\[-.1in]
v_1 & 1 &\Yfund & \Yfund \\
v_2 & \Yfund &1& \Yfund \\
v_3 & \Yfund & \Yfund&1 \\
v_E & 1 &\Yfund & \Yfund \\
v_G & \Yfund &1& \Yfund \\
v_H & \Yfund & \Yfund&1 \\
w_N & 1 & \Yasymm & 1\\
\end{array}
\eqn{vsii}
\eeq

The $v_{1,2,3}$ fields generate the $d$ quark Yukawa interactions
as in the toy model of the previous section; they also play a role in the $u$
quark Yukawa interactions, along with the $Sp(6)_L$ dynamical superpotential.
The fields $v_{E,G,H}$ differ from $v_{1,2,3}$ in that they do not participate
in
$v^3$ interactions, and so can only generate masses for composites in real
representations, as in fig.~1; they also couple to different combinations of
the $s,d,t$ preons.  For example, forbidding a coupling of either $v_2$ and
$v_3$ to the preon $s_1$ (a constituent of $H_u^{(1)}$) protects the zero in
the Higgs matrix \refeq{mhiggs} from contributions pictured in fig.~3;
however,
$s_1$ is also a constituent of $G$, and so  $v_G$ must couple to it or else the
$G$ and $\bar G$ composities will remain massless.

We give here a crude numerical fit to data, purposefully not fine tuned.
We take the six
$v$ fields to be degenerate, with the following couplings:
\beq
&& v_1\left[ s_{33} s_{23} + s_{31} s_{22}+  s_{21} s_{32}\right] -0.4
(v_1^2/2)\left[a_2 + a_3\right] \cr
+&& v_2 t_1 t_3 -0.4 (v_2^2/2)\left[a_1 + a_3\right] \cr
+&& v_3 d_1 d_2 + (v_3^2/2)\left[0.3 a_1 +0.7 a_2\right] \cr
+&& 12  v_E\left[   s_{31}  s'_{22}  +  s_{33}s_{23} \right]   -0.9 
(v_E^2/2)
\left[ a_2 + a_3\right] \cr
+&& v_G\left[ 0.1 t_1 t_3 + s_1 s_{32} \right] -0.4(v_G^2/2)\left[a_1 +
a_3\right] \cr
+&& v_H\left[ d_1 d_2 + s_1 s'_{22}\right] -  0.3 (v_H^2/2)\left[  a_1+
a_2\right]
\eeq
The scalar VEVs are
\beq
{\langle S^{(2)}_1\rangle\over \Lambda }= -0.05 {\Lambda\over4 \pi\mu}\ ,\qquad
{\langle ({S'}_2^{(2)})\rangle\over \Lambda }=- 0.03 {\Lambda\over4 
\pi\mu}\nonumber \\
{\langle S_2^{(3)}\rangle\over \Lambda }= 0.03 {\Lambda\over4 \pi\mu}\ ,\qquad
{\langle S_3^{(3)}\rangle\over \Lambda }= 10  {\Lambda\over4 \pi\mu}\ ,
\eeq
with
\beq
{\Lambda\over\mu}= {1\over 2}\ .
\eeq
With these parameters one finds the quark and lepton mass ratios (at the scale
$\Lambda$, which will be high),
\beq
&m_t/m_c= 410,\qquad &m_c/m_u = 300,\nonumber\\
&m_b/m_s=24, \qquad  &m_s/m_d = 34,\nonumber\\
&m_\tau/m_\mu=15, \qquad &m_\mu/m_e=1700\nonumber\\ 
&m_b/m_\tau=1\ .\qquad &
\eeq
The  CKM matrix (again computed at the scale $\Lambda$) is
\beq
V_{CKM} = \left(\matrix{
0.98 & 0.16 & 0.07\cr
-0.17 & 0.97 & 0.17 \cr
-0.05 & -0.18 & 0.98\cr
}\right)
\eeq

Aside from displaying sophisticated flavor structure, we consider this
model's interesting features to include a dynamically generated top mass,
as well as the  complex mixing between SM particles and vectorlike fields.
The model also displays the pitfalls generic to theories of composite quarks
and leptons that do not possess a baryon number symmetry, even if they
are invariant under $B-L$ parity: in this model,  dimension five proton decay
operators are generated upon integrating out the massive $G$ field.
We have not examined these operators in detail, but the proton lifetime is
expected  to be shorter than observed for any value of the compositeness scale
below the Planck scale.

\subsection{A realistic model with fundamental Higgs fields}

By taking the Higgs fields to be fundamental instead of composite, we can
simultaneously simplify the model and make it agree quantitatively with the
observed world.
In this section, we present an explicit example of such a model which
successfully fits all masses and
mixing angles of the SM.
In this model the three generations of matter fields are composites of two
strong $Sp(6)$ groups which confine at a scale $\Lambda$ as described in
\S2.
An interesting feature of this model is that baryon number is preserved
exactly,
thus preventing dangerous proton decay and allowing the compositeness
scale to be low enough that there are experimentally testable consequences.

\subsubsection{Field content}

We take the preons of the ultraviolet to transform under
the Standard Model gauge group with well-defined baryon and lepton number.

\beq
\begin{array}{c|cc|ccrrr}
{\rm preon} & Sp(6)_1 & Sp(6)_2 & SU(3) & SU(2) & U(1)_Y & U(1)_B &
U(1)_L \\[.1in] \hline&&&&&&&\\[-.1in]
a_1 & \Yasymm &1 & 1 & 1 & 0 & 0 & 0 \\
p_1 & \Yfund &1 & 3 & 1 & 1/6 & -1/6 & -1/2 \\
q_1 & \Yfund & 1& 1 & 2 & 0 & 1/2 & 1/2 \\
r_1 & \Yfund &1 & 1 & 1 & -1/2 & -1/2 & 1/2 \\[.1in] \hline&&&&&&&\\[-.1in]
a_2 &1 & \Yasymm & 1 & 1 & 0 & 0 & 0 \\
p_2 &1 & \Yfund & \mybar{3} & 1 & -1/6 & 1/6 & 1/2 \\
q_2 &1 & \Yfund & 1 & 1 & 1/2 & -1/2 & -1/2 \\
r_2 &1 & \Yfund & 1 & 1 & -1/2 & -1/2 & -1/2 \\
s_2 & 1& \Yfund & 1 & 1 & 1/2 & 1/2 & -1/2 \\
\end{array}
\eqn{preons}
\eeq

After confinement of the two $Sp(6)$ groups we obtain three generations of
composite quarks and leptons,
distinguished by the number of $a$ constituents.
The Standard Model quantum numbers of these composites are

\beq
\begin{array}{c|ccrrr}
{\rm composite} & SU(3) & SU(2) & U(1)_Y & U(1)_B & U(1)_L 
\\[.1in]\hline&&&&&\\[-.1in]
Q =p_1 q_1 & 3 & 2 & 1/6 & 1/3 & 0 \\
\mybar U =p_2 r_2 & \mybar{3} & 1 & -2/3 & -1/3 & 0 \\
\mybar D =p_2 q_2 & \mybar{3} & 1 & 1/3 & -1/3 & 0 \\
L =q_1 r_1 & 1 & 2 & -1/2 & 0 & 1 \\
\mybar E =q_2 s_2 & 1 & 1 & 1 & 0 & -1 \\
\mybar N =r_2 s_2 & 1 & 1 & 0 & 0 & -1 \\
G =p_2 p_2 & 3 & 1 & -1/3 & 1/3 & 1 \\
\mybar{G} =p_1 p_1 & \mybar{3} & 1 & 1/3 & -1/3 & -1 \\
R =p_2 s_2 & \mybar{3} & 1 & 1/3 & 2/3 & 0 \\
\mybar{R} =p_1 r_1 & 3 & 1 & -1/3 & -2/3 & 0 \\
S =q_1 q_1 & 1 & 1 & 0 & 1 & 1 \\
\mybar{S} =q_2 r_2 & 1 & 1 & 0 & -1 & -1 \\
\end{array}
\eqn{composites}
\eeq
where again we have only shown only one of three composite families;
the remaining families contain one or two $a$ preons.

In addition to the three SM families,
there are three generations of right handed neutrinos, as well as exotics
in real representations of both the Standard Model
gauge group and baryon and lepton number. As advertised, the SM Higgs
fields don't appear in the composite spectrum, they are added as fundamental
fields.

At this stage there are no couplings of the composites to the Higgs fields,
thus there are no SM Yukawa couplings. The dynamically generated
superpotential \refeq{Sp6superpot} of the confining $Sp(6)$ groups only
couples composites of the same strong group.
Since all of these couplings involve exotic fields which will be shown to
obtain large masses, these dynamical superpotential terms are irrelevant
for the infrared theory, and we will
not be concerned with them any further.

\subsubsection{Yukawa couplings}

In order to generate the Yukawa couplings we need couple
the preons of the two strong groups to the Higgs doublets. This is achieved by
introducing the following additional fields, all of which are taken to have
masses above the confinement scale.

\beq
\begin{array}{c|cc|ccrrr}
{\rm field} & Sp(6)_1 & Sp(6)_2 & SU(3) & SU(2) & U(1)_Y & U(1)_B &
U(1)_L \\[.1in]\hline&&&&&&&\\[-.1in]
v_u & \Yfund & \Yfund & 1 & 1 & 0 & 0 & 0 \\
v_d & \Yfund & \Yfund & 1 & 1 & 0 & 0 & 0 \\
v_s & \Yfund & \Yfund & 1 & 2 & -1/2 & 0 & 0\\
\mybar v_s & \Yfund & \Yfund & 1 & 2 & 1/2 & 0 & 0\\
w_u & \Yfund & 1 & 1 & 1 & 1/2 & 1/2 & 1/2 \\
\mybar w_u & \Yfund & 1 & 1 & 1 & -1/2 & -1/2 & -1/2 \\
w_d & \Yfund & 1 & 1 & 1 & -1/2 & 1/2 & 1/2 \\
\mybar w_d & \Yfund & 1 & 1 & 1 & 1/2 & -1/2 & -1/2 \\
\end{array}
\eqn{vs}
\eeq

The gauge symmetries not only allow masses for the new fields but also
renormalizable couplings to the preons of table \refeq{preons}.  We assume
a perturbative superpotential of the form
\beq
\eqn{Wtree}
\begin{array}{ll}
W_{tree}\!\!&=\frac{1}{2}(\mu-\alpha_u (a_1 + a_2)) v_u^2
        + \frac{1}{2}(\mu-\alpha_d (a_1 + a_2)) v_d^2
        + (\mu-\alpha_s (a_1 + a_2)) \mybar v_s v_s
\\
        &+\mu w_u \mybar{w}_u
        + \mu w_d \mybar{w}_d
        + \beta_1 v_u (p_1 p_2 + r_1 s_2)
        + \beta_2 v_d (p_1 p_2 +  r_1 s_2)
        + \beta_3 v_s q_1 q_2
\\
      &+\beta_4\mybar v_s q_1 r_2+\beta_5 v_u w_u r_2 + \beta_6 v_u w_dq_2
        + \beta_7 v_d w_d q_2
        + \beta_8 \mybar{w}_u H_u q_1 + \beta_9 \mybar{w}_d H_d q_1
\end{array}
\eeq
These couplings communicate between the different sectors and
generate masses and Yukawa couplings below the mass scales of the
heavy fields and confinement. The superpotential \eq{Wtree} is not the
most general allowed by the symmetries; to simplify the analysis, we have
identified the masses of the $v$ and $w$ fields, and left out a few couplings.
Superpotentials are well known to be non-generic, thus leaving out terms
which are allowed by symmetries is natural in a supersymmetric theory.
We will find that in this slightly simplified version of the model we can
get very close to fitting all the masses and angles of the SM with only
a few effective parameters, the theory with a generic superpotential whose
analysis we do not describe here, has more free parameters entering the Yukawa
matrices and can be made fully realistic.

To understand the origin of the SM Yukawa couplings we
follow the procedure described in the toy model in \S4:
first integrate out the massive $v$ and $w$ fields at the scale $\mu$ and
expand the
effective superpotential to dimension 4 in the preon fields $p,q,r,s$,
and to dimension 2 in each of the $a$ preons.
As in the toy model, below the mass of the $v$ and $w$'s
the gauge couplings of the $Sp(6)$
groups evolve rather quickly and get strong at scale $\Lambda$.
Preons are confined into composites, and the superpotential is mapped onto
an effective superpotential for the composites which contains mass terms
for the exotics (Fig 4.) and the desired Yukawa
couplings for SM quarks and leptons (Fig 5.)
\begin{figure}[t]
\centerline{\epsfxsize=3 in \epsfbox{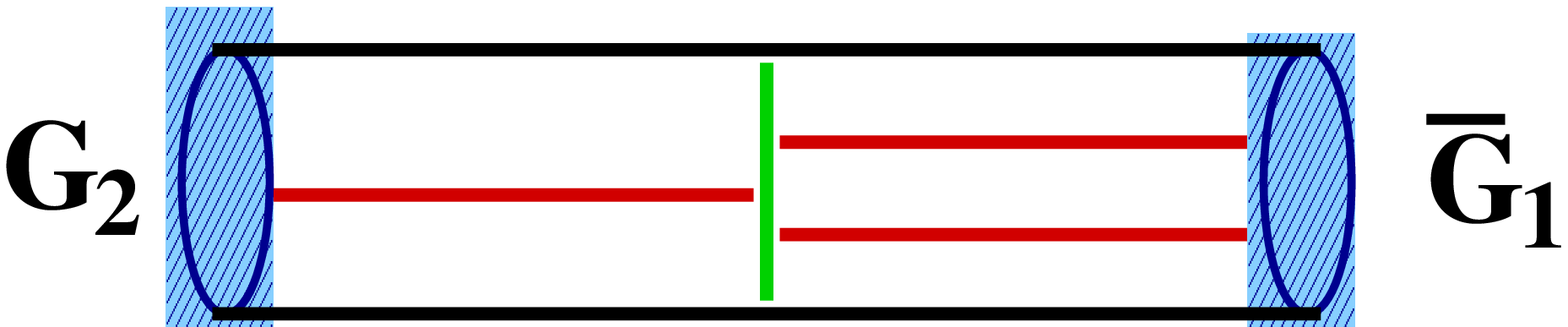}}
\vskip.2in
\noindent
Fig 4. {\it Contribution to the mass of $G$, $\mybar G$ in the effective
superpotential \eq{compo}. The internal (green) line is the
propagator of a heavy $v_u$ or $v_d$ field and the (black) lines at the top and
bottom of
the diagram are massless preons which carry SM quantum numbers
and are bound into composites together with SM gauge neutral preons
(red) by the confining dynamics (blue).  The mass generation for the
$R$, $\mybar R$ fields proceeds through similar diagrams, while the $S$,
$\mybar S$ mass arises from $v_s$, $\mybar v_s$ exchange. }
\vskip .2in
\end{figure}
\begin{figure}[t]
\centerline{\epsfxsize=3 in \epsfbox{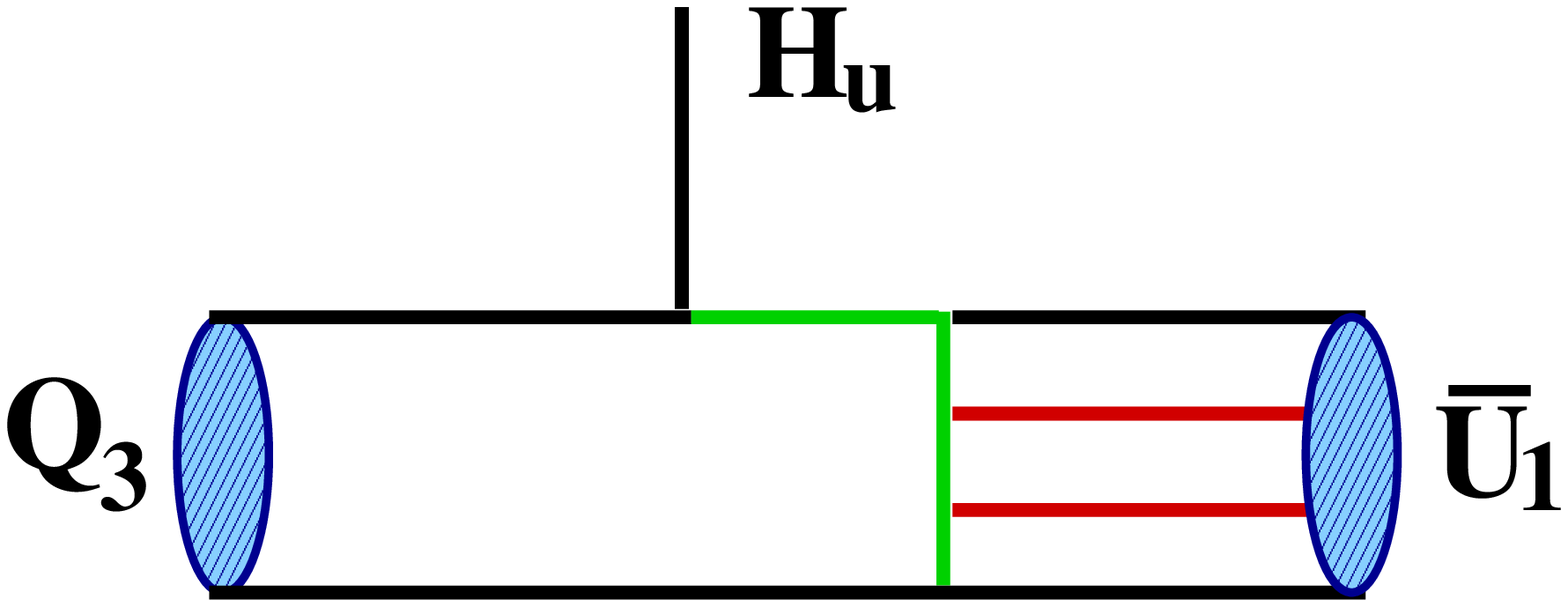}}
\vskip.2in
\noindent
Fig 5. {\it Contribution to SM Yukawa coupling $Y^u_{31}$. Internal (green)
lines correspond to massive $v$ and $w$ propagators, black (red)
lines are preons which carry (do not carry) SM quantum numbers and
are bound into composites by the nonperturbative strong dynamics (blue).
The Higgs fields in this model are fundamental and have perturbative
couplings to preons.}
\vskip .2in
\end{figure}
\beq
\eqn{compo}
W_{eff}&=&M^G_{ij} G_i \mybar{G}_j + M^R_{ij} R_i \mybar{R}_j  +
          M^S_{ij} S_i \mybar{S}_j\nonumber\\
         &&+Y^u_{ij} Q_i \mybar U_j H_u + Y^d_{ij} Q_i \mybar D_j H_d +
              Y^l_{ij} L_i \mybar E_j H_d +Y^n_{ij} L_i \mybar N_j H_u\ .
\eeq
The masses and Yukawa couplings of the effective theory are readily computed in
terms of the small parameters
\beq
\eu = \alpha_u {\Lambda \over \mu}\ ,
\quad \ed = \alpha_d {\Lambda \over \mu}\ ,
\quad \es = \alpha_s {\Lambda \over \mu}\ .
\eeq
The mass matrices for the exotic fields, $M^G,M^G$ and $M^S$ are
\beq
\eqn{exotmass}
M^{G}=M^R &=& {\Lambda^2\over \mu}\left[
 \beta_1^2 \left(\matrix{
        6\eu^4 & 3\eu^3 & \eu^2 \cr
       3\eu^3 & 2 \eu^2 & \eu\cr
       \eu^2& \eu & 1  \cr  }\right)
+ \beta_2^2 \left(\matrix{
        6\ed^4 & 3\ed^3 & \ed^2 \cr
       3\ed^3 & 2 \ed^2 & \ed\cr
       \ed^2& \ed & 1  \cr  }\right)\right]\ , \nonumber \\
M^{S} &=& {\Lambda^2\over \mu}\beta_3\beta_4 \left(\matrix{
        6\es^4 & 3\es^3 & \es^2 \cr
       3\es^3 & 2 \es^2 & \es\cr
       \es^2& \es & 1  \cr  }\right)\ .
 \eeq
When these fields are integrated out of the theory, they will induce  dimension
5 operators with interesting flavor structure into the effective 
superpotential. 
For example, integrating out the $G$  and $R$ fields respectively
gives rise to operators such as
\beq
{1\over M^G}QL
\mybar U\mybar E \quad {\rm and} \quad {1\over M^G}QQ\mybar D\mybar U\ .
\eqn{egop}
\eeq
The flavor structure of these operators is intricate due to the hierarchical
nature of the mass matrices \refeq{exotmass} and the structure of the instanton
induced couplings between the exotic and ordinary fields.

The Yukawa matrices for the quarks and leptons (renormalized at the
compositeness scale)  are given by
\beq
\eqn{mod2yuk}
Y^u\!=\lambda_u\!\left(\matrix{
        6\eu^4 & 3\eu^3 & \eu^2 \cr
       3\eu^3 & 2 \eu^2 & \eu\cr
       \eu^2& \eu & 1  \cr  }\right)\!,&&\hspace{-.2in}
Y^d\!=\lambda_d\!\left[\!\left(\matrix{
        6\ed^4 & 3\ed^3 & \ed^2 \cr
       3\ed^3 & 2 \ed^2 & \ed\cr
       \ed^2& \ed & 1  \cr  }\right)
       \!+ r \left(\matrix{
        6\eu^4 & 3\eu^3 & \eu^2 \cr
       3\eu^3 & 2 \eu^2 & \eu\cr
       \eu^2& \eu & 1  \cr  }\right)\!\right] \\
Y^n\!=\lambda_n\!\left(\matrix{
        6\eu^4 & 3\eu^3 & \eu^2 \cr
       3\eu^3 & 2 \eu^2 & \eu\cr
       \eu^2& \eu & 1  \cr  }\right),&&\hspace{-.2in}
Y^l\!=\lambda_l\left[\!\left(\matrix{
        6\ed^4 & 3\ed^3 & \ed^2 \cr
       3\ed^3 & 2 \ed^2 & \ed\cr
       \ed^2& \ed & 1  \cr  }\right)
       \!+ s \left(\matrix{
        6\eu^4 & 3\eu^3 & \eu^2 \cr
       3\eu^3 & 2 \eu^2 & \eu\cr
\eu^2& \eu & 1  \cr  }\right)\!\right]
\eeq
The $\lambda$ factors contain an overall $(\Lambda/\mu)^2$ times products of
the $\beta$ couplings from the ultraviolet superpotential \refeq{Wtree}; $r$
and $s$ are functions of the couplings as well.
 Note that in order for the top Yukawa coupling to be sufficiently large,
$\Lambda \sim \mu$ or  large $\beta$ couplings are required.

\subsubsection{Numerical predictions for masses and angles}

From \refeq{mod2yuk}, one might expect to obtain
predictions for four of the seven
quark mass ratios and mixing angles because the mass ratios and
angles only seem to depend on the three parameters $\eu,\ed$ and $r$. However,
to predict the quark and lepton masses and mixing angles
at the weak scale one needs to calculate the renormalization group evolution
of the couplings from the scale of confinement to the masses of
the quarks. The running depends sensitively on the value
of the top Yukawa coupling as well as the scale of confinement.
At one loop these contributions can be summarized in a single parameter
\cite{DHRBBO} so that we only lose one prediction.

A crude fit to $m_d/m_s,m_s/m_b,V_{cb},m_c/m_t$ yields the following
values for the parameters
\beq
\eu\sim0.045, \quad \ed\sim 0.22, \quad r\sim4.8\ .
\eeq
These values can then be used to predict
\beq
m_c/m_u=490, \quad V_{us}= 0.25, \quad {V_{ub}\over V_{cb}}= 0.17 \ .
\eeq

Whereas these numbers are not in complete agreement with experiment,
they are nevertheless  encouragingly close. This suggests that the textures
obtained in \eq{mod2yuk} might be interesting to study in their own
right.
In the context of this model, the precise numbers for the predictions
above should not be taken too seriously because wave function renormalization
for the composites arising from the unknown K\"ahler potential can alter these
predictions by factors presumed to be ${\cal O}(1)$.

In the charged lepton sector the model has one more parameter $s$ to
describe mass ratios. Fitting
 to $m_e/m_\mu$  yields a prediction for $m_\mu/m_\tau$ that is a
factor $1.4$ too large.
The fit can be improved while avoiding fine-tuning by including an additional
set of
intermediate fields $w_l$, $\mybar{w}_l$ which couple to the constituents of
the charged leptons.

Neutrino masses can be treated in the same manner as discussed above in
\S5.1.5. We introduce
an extra massive field $w_N$ with lepton number one which transforms
as an antisymmetric tensor of the  $Sp(6)_2$ group and include the
following superpotential couplings
\beq
W_N=\frac{1}{2}(\mu-\alpha_N a_2) w_N^2  + w_N r_2 s_2 \ .
\eeq
This interaction gives rise to the Majorana mass matrix \refeq{mnr} for the
right-handed neutrinos, and after they are integrated out of the theory,
the left-handed neutrinos develop hierarchical masses with
mass ratios of order $\eu^4/\eN^2$ through the seesaw mechanism.
Since the overal size $\lambda_n$
of the Dirac neutrino masses in  \eq{mod2yuk} is unconstrained,
neutrino mass bounds do not place any constraint on the scale of
compositeness.

\subsubsection{Comments}

In summary, this model based on $Sp(6)^2$ is successful at explaining the
existence of three families and in reproducing the observed flavor
properties of quarks and leptons. A great virtue of the model is that it
possesses a baryon symmetry, so that the compositeness scale need not be high.
Constraints on the scale of compositeness will therefore come from  flavor
changing operators, such as lepton flavor violating  operators in the 
superpotential
\eq{egop} or in the K\"ahler potential. An investigation of such effects
will be pursued elsewhere; here we simply stress that compositeness
effects may show up in exotic flavor violating processes.

\section{Conclusions and Outlook}

The new dynamical approach to the flavor problem that we are advocating
has been shown to be capable of reproducing the fermion masses and
mixing angles seen in Nature, as well as explaining the replication of 
families.
We emphasize that we were able to demonstrate,  entirely in the context of
renormalizable field theories, how compositeness for quarks and leptons
can give rise to realistic flavor structure at low energies.  By not  resorting
to ``Planck slop'' or other high dimension operators, we  have not hidden
any of the dynamics responsible for flavor.

In the examples we presented, the three generations of quarks and leptons are
composites of new strong interactions.
The Yukawa couplings arise from a combination of perturbative
dynamics above the scale where the new gauge interactions get strong and
confinement. We obtain new textures of Yukawa
matrices \refeq{toyyuk} without zeros (but with calculable combinatoric 
factors)
which can be predictive and sufficiently rich to be realistic.
The more realistic models we considered in \S5 are not unifiable, but as
demonstrated by our toy model of \S4,  the flavor mechanism we are
proposing here is at least in principle compatible with unification.

There are a number of avenues to explore from here:

\begin{enumerate}
\item {\it Other strong groups.}
In this paper we have focused on the gauge group $Sp(6)$ with an antisymmetric
tensor matter field to generate three generations.
There are many other models which might be useful for generating
composite generations.
A promising example is supersymmetric $SU(N_c)$ gauge theory with an
adjoint matter field $A$ and $N_f$ flavors $Q+\mybar Q$. Gauge invariants of
this theory include the composite operators $M_j=QA^j\mybar Q$ for
$j=0,1,2,\cdots$.
The dynamics of these theories is not as well understood, but with
an added tree level superpotential $W_{tree}=\rm tr A^{k+1}$ the theory
has a dual description in terms of an $SU(kN_c-N_f)$ gauge group~\cite{KSS}, 
and
is believed to confine and generate $k$ generations of composites $M_j$
for the special case $kN_c-N_f=1$. Other models can be constructed using
$SO, SU, Sp$ groups with various tensor matter fields \cite{CSS,otherKS}.

\item {\it SUSY breaking.}
The models presented in this paper are incomplete as they
do not address the issue of SUSY breaking and electroweak
symmetry breaking. It is desirable to have the nonperturbative scale of flavor
physics simultaneously explain the electroweak hierarchy.  However, relating
flavor physics and SUSY breaking runs the risk of generating large flavor
changing neutral currents through nondegeneracy of squark and slepton masses.
To avoid large flavor changing effects, it seems worthwhile to see if the 
strong
flavor dynamics we envision could trigger SUSY breaking  in a low energy
gauge-mediated model ~\cite{gaugemed}. An alternative would be to try to
realize our flavor mechanism within the Effective SUSY scenario \cite{EffSUSY},
in which dangerous flavor changing processes are suppressed by having first
and second family sparticles be much heavier than those of the third family.
 A technical problem is to
find a model which breaks SUSY while preserving a large enough
non-Abelian global symmetry into which the SM gauge group can be imbedded.
Conventional models of dynamical SUSY breaking are constructed by
lifting all classical flat directions with a tree level superpotential in a
theory in which quantum effects push the vacuum away from the
origin, resulting in non-zero vacuum energy and SUSY breaking~\cite{witek}.
These models are unsatisfactory for our purposes since both the
tree level  superpotential and the VEVs break the desired global symmetries.
Recently, models of dynamical SUSY
breaking with sufficiently large global symmetries have been
constructed~\cite{recgaugemed,ann}; a common feature of these models
are classical flat directions which are only lifted by quantum dynamics.

\item {\it Phenomenology.} It seems worth pursuing the phenomenology
of flavor changing interactions that result from compositeness, such as those
discussed briefly in the previous section. It is not clear, however, how much
can be said in a model without SUSY breaking.  Another feature of
phenomenological interest in the models we have described is the ubiquity
of neutral moduli such as the $T_k={\rm tr} A^k$ composite fields.
While they can be expected to develop mass when SUSY is broken, there
may be many such fields, with flavor dependent couplings, which could in
principle mediate detectable long range forces \cite{forces} .

\end{enumerate}

\section{Acknowledgements}
We are grateful to Ann Nelson for useful
conversations. D.B.K and F.L. are  supported in part by DOE grant
DOE-ER-40561, and NSF Presidential Young Investigator award
PHY-9057135.
M.S. is supported by the U.S.
Department of Energy under grant \#DE-FG02-91ER40676.


\begin{thebibliography}{99}
\bibitem{susyref}
H.P. Nilles, \PR{110}{1}{1984}.

\bibitem{Seib}
N. Seiberg, \PRD{49}{6857}{1994}, hep-th/9402044;
\NPB{435}{129}{1995}, hep-th/9411149.

\bibitem{exactres}
{\it for reviews see e.g.}
K. Intriligator and N. Seiberg, {\it Nucl. Phys. Proc. Suppl.} {\bf 45BC}, 1 
(1996), 
hep-th/9509066;
M.E. Peskin, hep-th/9702094.

\bibitem{copapers}
M. Strassler, \PLB{376}{119}{1996}, hep-ph/9510342;
A. Nelson and M. Strassler, hep-ph/9607362;
M.A. Luty and R.N. Mohapatra, \PLB{396}{161}{1997}, hep-ph/9611343.

\bibitem{CSS}
C. Csaki, M. Schmaltz, W. Skiba, \PRL{78}{799}{1997}, hep-th/9610139;
hep-th/9612207.

\bibitem{CHM}
C.D. Carone, L.J. Hall and T. Moroi, hep-ph/9705383.

\bibitem{Cho}
P. Cho and P. Kraus, \PRD{54}{7640}{1996}, hep-th/9607200.

\bibitem{oursp}
C. Cs\'aki, M. Schmaltz and W. Skiba, \NPB{487}{128}{1997}, hep-th/9607210.

\bibitem{FN}
C.D. Froggatt and H.B. Nielsen, \NPB{147}{277}{1979}.

\bibitem{abelflav}
Y. Nir and N. Seiberg,  \PLB{309}{337}{1993}, hep-ph/9304307;
M. Leurer, Y. Nir and N. Seiberg, \NPB{398}{319}{1993}, hep-ph/9212278;
M. Leurer, Y. Nir and N. Seiberg, \NPB{420}{468}{1994}, hep-ph/9310320.

\bibitem{nonabelflav}
P. Pouliot and N. Seiberg, \PLB{318}{169}{1993}, hep-ph/9308363;
D.B. Kaplan and M. Schmaltz, \PRD{49}{3741}{1994}, hep-ph/9311281;
P.H. Frampton and O.C.W. Kong \PRD{53}{2293}{1996}, hep-ph/9511343;
R. Barbieri, G. Dvali and L.J. Hall, \PLB{377}{76}{1996}, hep-ph/9512388;
K.S. Babu and S.M. Barr; \PLB{387}{87}{1996}; hep-ph/9606384.

\bibitem{EffSUSY}
A. Cohen, D. Kaplan and A. Nelson,  \PLB{388}{588}{1996}, hep-ph/9607394.

\bibitem{trinification}
S.L. Glashow, Published in Providence Grand Unification 88 (1984).

\bibitem{DHRBBO}
S. Dimopoulos, L.J. Hall and S. Raby, \PRL{68}{1984}{1992};
V. Barger, M.S. Berger and P. Ohmann, \PRD{47}{1093}{1993},
hep-ph/9209232; \PRD{47}{2038}{1993}, hep-ph/9210260.

\bibitem{KSS}
D. Kutasov, \PLB{351}{230}{1995}, hep-th/9503086;
D. Kutasov and A. Schwimmer, \PLB{354}{315}{1995}, hep-th/9505004;
D. Kutasov, A. Schwimmer, and N. Seiberg, \NPB{459}{455}{1996}, hep-th/9510222.

\bibitem{otherKS}
K. Intriligator, \NPB{448}{187}{1995}, hep-th/9505051;
K. Intriligator, R. Leigh, and M. Strassler, \NPB{456}{567}{1995},
hep-th/9506148;
J. Brodie and M. Strassler, hep-th/9611197.

\bibitem{gaugemed}
M. Dine and A. Nelson, \PRD{48}{1277}{1993}, hep-ph/9303230;
M. Dine, A. Nelson and Y. Shirman, \PRD{51}{1362}{1995}, hep-ph/9408384;
M. Dine, A. Nelson, Y. Nir and Y. Shirman, \PRD{53}{2658}{1996},
hep-ph/9507378;
L. Randall, hep-ph/9612426.

\bibitem{witek}
W. Skiba, hep-th/9703159; {\it and references therein.}

\bibitem{recgaugemed}
K. Intriligator and S. Thomas, \NPB{473}{121}{1996}, hep-th/9603158;
E. Poppitz and S.P. Trivedi, \PRD{55}{5508}{1997}, hep-ph/9609529;
N. Arkani-Hamed, J. March-Russell and H. Murayama, hep-ph/9701286;
K.I. Izawa, Y. Nomura, K. Tobe and T. Yanagida, hep-ph/9705228;
H. Murayama, hep-ph/9705271.

\bibitem{ann}
A. Nelson, {\it private communication}.

\bibitem{forces}
S. Dimopoulos and G.F. Giudice, \PLB{379}{105}{1996}, hep-ph/9602350.

\end{thebibliography}
\end{document}